\documentclass[pra,showpacs,aps,amssymb,amsmath,nofootinbib,notitlepage,superscriptaddress,twocolumn,longbibliography]{revtex4-1}

\usepackage{amsfonts}
\usepackage{graphicx,graphics,epsfig,times,bm,bbm,mathrsfs} 
\usepackage[normalem]{ulem}
\usepackage{subfigure}
\usepackage[pdfstartview=FitH]{hyperref}
\hypersetup{
    colorlinks=true,       
    linkcolor=cyan,          
    citecolor=magenta,        
    filecolor=magenta,      
    urlcolor=cyan,           
    runcolor=cyan
}
\usepackage{upgreek}

\newcommand{\beq}{\begin{equation}}
\newcommand{\eeq}{\end{equation}}
\newcommand{\beqnn}{\begin{equation*}}
\newcommand{\eeqnn}{\end{equation*}}
\newcommand{\beann}{\begin{eqnarray*}}
\newcommand{\eeann}{\end{eqnarray*}}

\newcommand{\bes} {\begin{subequations}}
\newcommand{\ees} {\end{subequations}}
\newcommand{\bea} {\begin{eqnarray}}
\newcommand{\eea} {\end{eqnarray}}

\newcommand{\ignore}[1]{}

\newcommand{\ket}[1]{ | #1\rangle}

\newcommand{\run}{\mathrm{run}}

\begin{document}

\title{Optimally Stopped Variational Quantum Algorithms}

\author{Walter Vinci}

\affiliation{Department of Electrical Engineering, University of Southern California, Los Angeles, California 90089, USA}

\affiliation{Department of Physics and Astronomy, University of Southern California, Los Angeles, California 90089, USA}

\affiliation{Center for Quantum Information Science \& Technology, University of Southern California, Los Angeles, California 90089, USA}

\author{Alireza Shabani}

\affiliation{Qulab Inc., Los Angeles, California 90049, USA}

\date{\today}

\begin{abstract}
Quantum processors promise a paradigm shift in high-performance computing which needs to be assessed by accurate benchmarking measures.
In this work, we introduce a new benchmark for variational quantum algorithm (VQA), recently proposed as a heuristic algorithm for small-scale quantum processors. In VQA, a classical optimization algorithm guides the quantum dynamics of the processor to yield the best solution for a given problem. A complete assessment of scalability and competitiveness of VQA should take into account both the quality and the time of dynamics optimization. The method of optimal stopping, employed here, provides such an assessment by explicitly including time as a cost factor. Here we showcase this measure for benchmarking VQA as a solver for some quadratic unconstrained binary optimization. Moreover we show that a better choice for the cost function of the classical routine can significantly improve the performance of the VQA algorithm and even improving it's scaling properties.

\end{abstract}

\maketitle

\section{Introduction} Variational Quantum Algorithm (VQA) has recently been introduced as a promising approach for computation on small-scale quantum processors~\cite{Peruzzo:14,Farhi,McClean:16,McClean:16-2,Omalley:15,wecker,Yang:17,Wang:17,Kandala:17}. It was originally introduced as an ab-initio solver for electronic structure problem~\cite{Peruzzo:14}, a generalization of unitary coupled-cluster method \cite{Bartlett:07}, with applications in chemistry and material science. It was later applied for optimizing classical functions ~\cite{Farhi}. VQA is often referred to as Variational Quantum Eigensolver for quantum chemistry \cite{Peruzzo:14,McClean:16,McClean:16-2,Omalley:15,Kandala:17}, or Quantum Approximate Optimization Algorithm for classical optimization applications \cite{Farhi,wecker,Yang:17,Wang:17}. In VQA, a computational task is encoded as an optimization problem over quantum states generated by a parametrized quantum dynamics. Computation is accomplished via a classical search over the space of dynamics-free parameters, for instance system-Hamiltonian parameters. Ref.~\cite{Yang:17} casts VQA as a closed-loop optimal quantum control problem, basically a quantum observable control problem \cite{Rabitz:06}. From computer science perspective, in VQA, instead of intelligently designing the gate sequence for a quantum algorithm, one finds a good approximation through a search process. This makes VQA a quantum counterpart to classical feedforward neural networks. 

Since VQA is a heuristic hybrid quantum-classical algorithm, it becomes important to benchmark it properly against classical algorithms. So far VQA studies have been focused on accuracy of computation ignoring the time of search for optimal quantum evolution ~\cite{Peruzzo:14,Farhi,McClean:16,McClean:16-2,Omalley:15,wecker,Yang:17,Wang:17}. For most applications, both accuracy and total time of computation are key factors to determine the performance of an algorithm. Therefore any proper benchmark should measure the performance based on a trade-off between time and accuracy. In this manuscript, we consider such a trade-off by considering optimal stopping cost~\cite{vinci2016optimally} as a benchmarking measure. 
We exemplify advantages of using optimal costs by solving some quadratic unconstrained binary optimization (QUBO). We also use optimal costs to define a convenient cost function for the classical optimization routine that has more weight on the low energy solutions. It turns out that this approach significantly improves the performance of the VQA algorithm in solving classical combinatorial problems, as compared to the common choice to optimize the expectation value of the problem Hamiltonian.  We begin in Sec.~\ref{sec:VQA} by reviewing the definition of VQA. In Sec.~\ref{sec:OPT}, we review optimal stopping approach to benchmarking optimization algorithms. In Sec.~\ref{sec:NUM}, we present numerical results and conclude in Sec.~\ref{sec:DIS}.

\section{Variational Quantum Algorithm}
\label{sec:VQA}

Consider an $N-$qubit quantum circuit and Hamiltonian $H(\{g_\alpha(t)\})=\sum_\alpha g_\alpha(t) H_\alpha $ with tunable parameters $g_\alpha(t)$. A typical quantum algorithm prescribes a sequence of gates, via turning variables $g_\alpha(t)$ on and off, making a series of unitary gates. In VQA, instead, one finds the right dynamics via a search over variables $g_\alpha(t)$ rather than designing them, as follows
\begin{enumerate}
\item Cast the computational problem as minimization of a function $E(|\Psi\rangle)$ such that the solution $|\Psi_S\rangle$ is the argument 
\begin{equation}
|\Psi_S\rangle=\arg \min_{|\Psi\rangle} E(|\Psi\rangle)\,.
\end{equation} 
For the electronic structure problem $E(|\Psi\rangle)=\langle\Psi|H|\Psi\rangle$ where $H$ is the system Hamiltonian under the Born-Oppenheimer approximation \cite{Peruzzo:14,McClean:16,McClean:16-2,Omalley:15,Kandala:17}. Similarly, for classical optimization problems, $H$ is an Ising Hamiltonian \cite{Farhi,wecker,Yang:17,Wang:17} encoded the same way as in quantum annealing \cite{Das:08,Albash:16,Neven:17}.
\item Initialize system in a trivial state, e.g. $|\Psi^{\rm in}\rangle=|0\rangle^{\otimes N}$, with some choice of parameters values $\{g_\alpha(t)\}$.
\item Solve the following optimization problem: 
\begin{eqnarray}
\{g_\alpha^*\}=\arg\min_{\{g_\alpha\}} E(|\Psi^{\rm out}(g_\alpha)\rangle)\label{main}\\
\mbox{where\hspace{0.1 in}} |\Psi^{\rm out}(g_\alpha)\rangle=U_T[g_\alpha(t)] |\Psi^{\rm in}\rangle\notag
\end{eqnarray}
where the circuit unitary evolution $U_T(g_\alpha(t))$ is generated by the Hamiltonian $H(\{g_\alpha(t)\})$ for a total time $T$.
\end{enumerate}
The (approximate) solution of the computational problem is therefore $|\Psi_S\rangle=|\Psi^{\rm out}(g^*_\alpha)\rangle$. The optimization problem (\ref{main}) is solved by running a classical optimization problem. Nelder-Mead simplex algorithm and quasi-Newton methods are two commonly used algorithms that we consider for our simulations. Any optimization algorithm solving (\ref{main}) iteratively uses the circuit evolution $U_T$ as a subroutine. Therefore the total time of quantum computation using VQA is given by the runtime of the quantum circuit subroutine times the number of calls to the quantum circuit required by the classical optimization (\ref{main}).   

Note that VQAs are commonly considered for circuit model quantum computation. However, its principle can be extended beyond circuit model dynamics, to dissipative systems as discussed in Ref.~\cite{Yang:17}, or for quantum annealing~\cite{Das:08,Albash:16,Neven:17,Lanting:17}. Next section is an introduction to the notion of optimal stopping which we employ to benchmark VQAs as well as to define a new cost function for the optimization of a VQA.

\section{Optimal Stopping background}
\label{sec:OPT}

In the previous section we have described a general setup for a VQA. In this paper, we are interested in using VQA for solving classical optimization, for which the function $E$ is a diagonal operator in the computational basis $z_i$. Moreover, the only acceptable solutions are the states of the computational basis itself. In this setup, a candidate solution is obtained by performing an additional measurement in the computational basis of the output state $|\Psi^{\rm out}(g^*_\alpha)\rangle$.

The fundamental assumption to use optimal stopping is that we can describe an optimization algorithm in terms of an intrinsic ``quality distribution" $\mathcal P(e)$ of the qualities $e$ of the outcomes. From now on we will mostly use the term ``energy" to indicate the quality of a solution. This setup is particularly appropriate for the VQA since the states of the computational basis are obtained after projective measurements of the computational basis $z_i$ on the final state obtained by the VQA:
\begin{equation}
\mathcal P(e_a) = \sum_{i | E(z_i) = e_a}|\langle z_i  | \Psi^{\rm out} \rangle  |^2\,,
\end{equation}
where $e_a$ is the set of unique energies for all configurations. The idea behind benchmarking via optimal stopping~\cite{vinci2016optimally} is to minimize a \emph{total cost} $C(t)$ as a function of the computation time $t$:
\beq
C(t) = E(t) + T(t)\,,
\label{eq:gen}
\eeq
where $E(t)$ is the quality of the best solution found at time $t$ (without loss of generality we assume that $E(t)$ is the value of the objective function, which we may think of as an energy, that defines the optimization problem) and $T(t)$ is a measure of the computational effort.  Typically, and this is the case for VQA, time is discretized in steps of $t_{\run}$, e.g. the iteration of the classical optimization routine.   The total cost $C(t)$  can thus be rewritten as:
 \beq
 C_n =  \min \{e_1,\dots,e_n\} +T_n = E_n+T_n\,.
\label{eq:expense}
 \eeq

The optimal total cost is then by definition the average (expected) cost obtained when following an optimal stopping rule:
\beq
C^* \equiv \langle C_{n^*} \rangle = \langle E_{n^*} \rangle + \langle T_{n^*} \rangle \equiv E^*+ T^*\,,
\label{eq:split}
\eeq
where the average is taken over several repeated optimally stopped sequences,  $E^*$ is the optimal energy and $T^*$ is the optimal computational effort. The optimal cost can be obtained following the optimal stopping rule~\cite{ferguson}:
\beq
n^* = \min\{n \ge 1: e_n \le C^*\}\,.
\label{eq:opt_n}
\eeq

\begin{figure*}[t]
\begin{center}
\subfigure[\, ]{\includegraphics[width=0.36\textwidth]{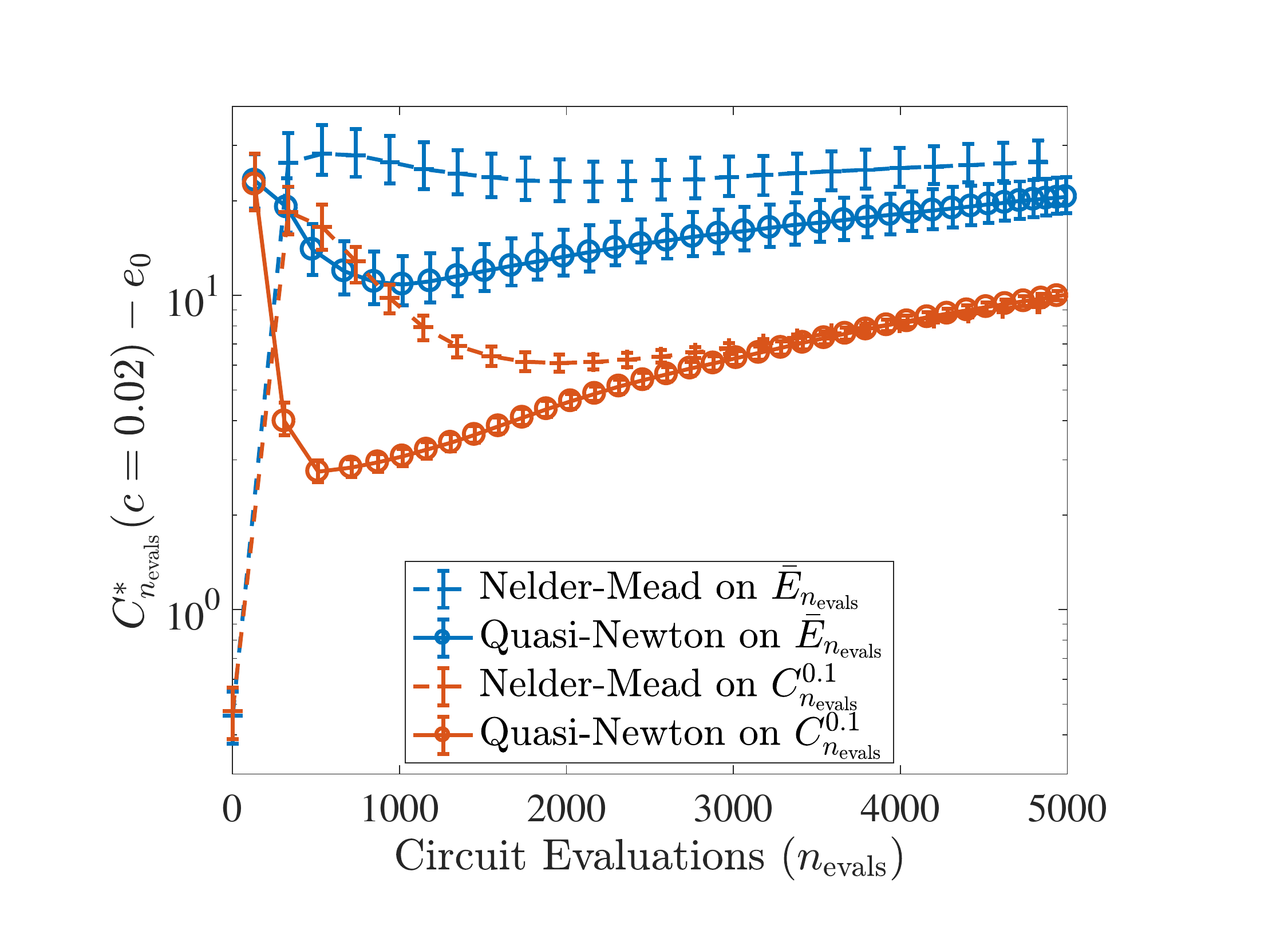}\label{fig:1a}} \hspace{-1cm}
\subfigure[\, ]{\includegraphics[width=0.36\textwidth]{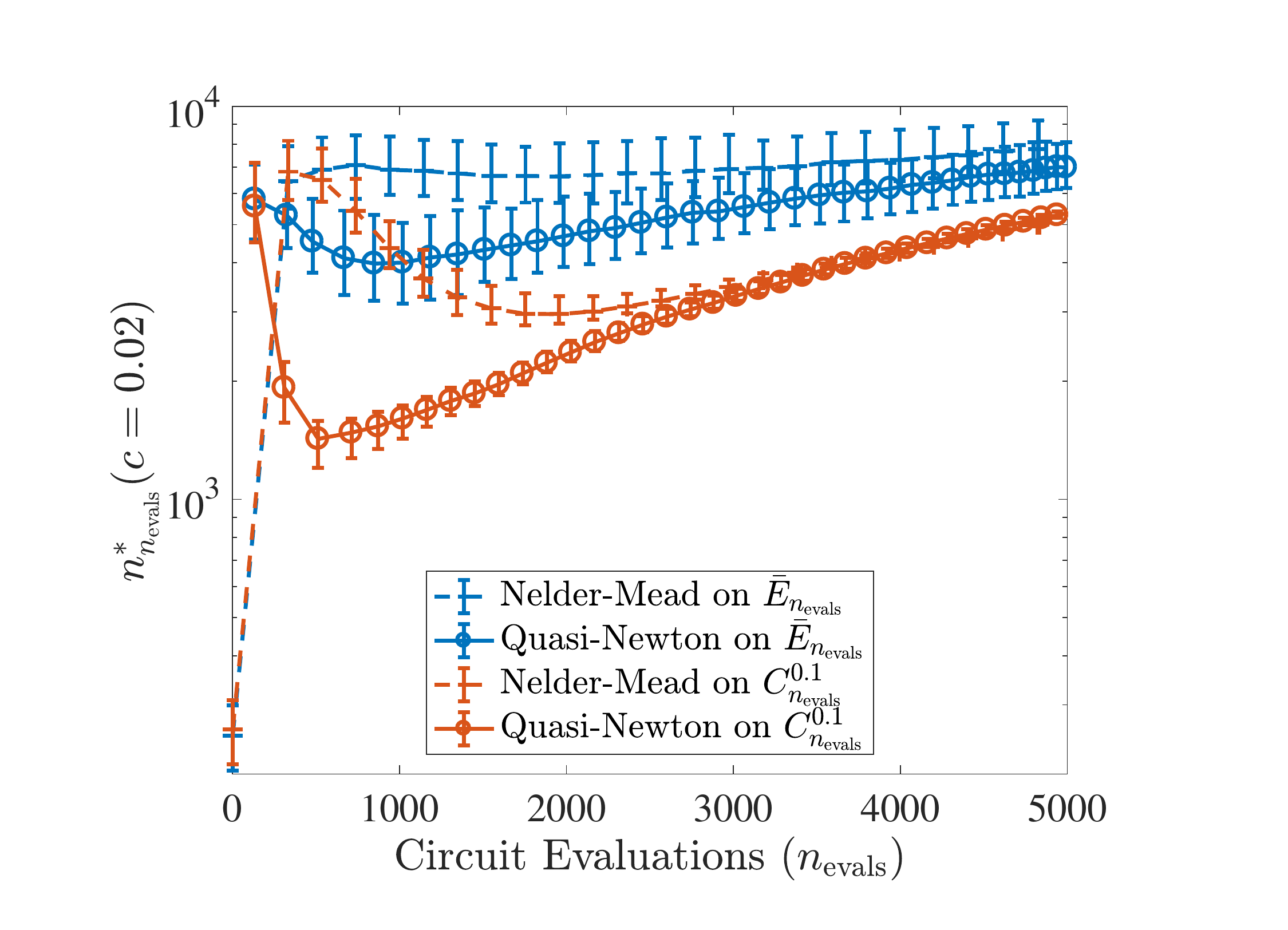}\label{fig:1b}} \hspace{-1cm}
\subfigure[\, ]{\includegraphics[width=0.36\textwidth]{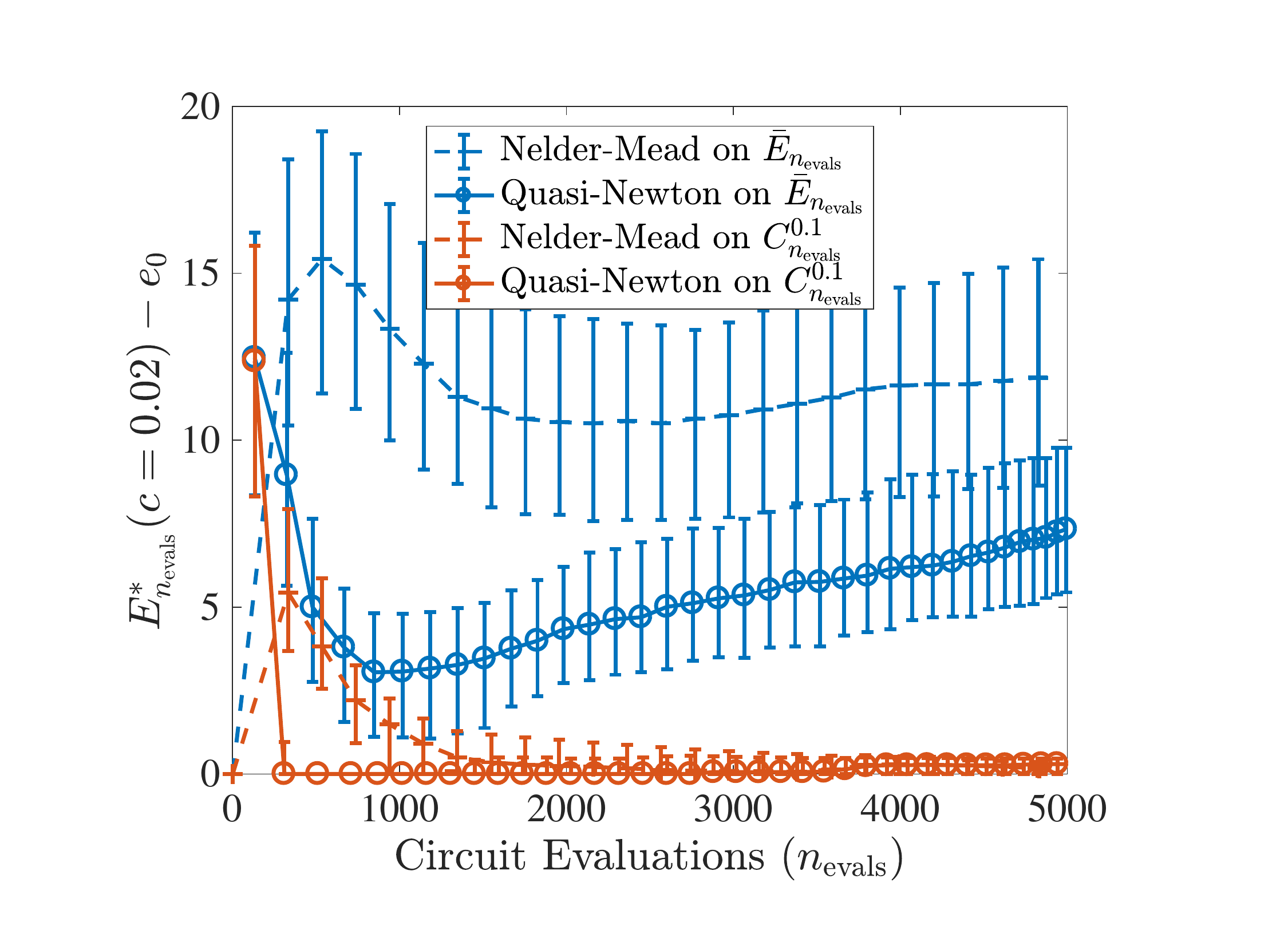}\label{fig:1c}}
\caption{Performance of VQA (depth-3 quantum circuit) on a randomly generated ensemble of 100 QUBO problems with $N=8$ as a function of the number of circuit evaluations ${n_{\rm evals}}$.  The solvers $\tt fminsearch$  and  $\tt fminunc$ are used for the classical optimization process that attempts to minimize either $C^{0.1}_{n_{\rm evals}}$ or $\bar E_{n_{\rm evals}}$. (a) Optimal cost $C^*_{n_{\rm evals}}(c)$. (b) Optimal number of circuit evaluations  $n^*_{n_{\rm evals}}(c)$. (c) Optimal energy $E^*_{n_{\rm evals}}(c)$. All quantities are  computed at $c=0.02$. Recall from Eq.~\ref{eq:split} that $C^*_{n_{\rm evals}}(c) = E^*_{n_{\rm evals}}(c) + c n^*_{n_{\rm evals}}(c)$.}
\label{fig:1}
\end{center}
\end{figure*}

We consider a special, but practically relevant case that can be solved analytically. We assume that the cost function is linear in time:
\beq
 C_n =  \min \{e_1,\dots,e_n\} +n c t_{\run}\,,
\label{eq:linexp}
 \eeq
where the parameter  $c$ is interpreted as the cost per unit of time that specifies the computational effort. Large $c$ favors short computations over good quality solutions, vice versa smaller $c$ favors obtaining good solutions over time. With a linear in time computational effort, the optimal total cost $C^*$  is the solution of the following \emph{optimality equation}~\cite{ferguson}:
\beq
 C^*(c)  \,\,:\,\, \sum_{a | e_{\rm g} \le e_a \le C^*(c)}(C^*(c)-e_a)\mathcal P(e_a)  =  c t_{\run}\,,
 \label{eq:opt_E}
\eeq
where $e_{\rm g}$ is the ground state energy.  Because of the optimal stopping rule Eq.~\ref{eq:opt_n}, $C^*(c)$ can be interpreted as an energy target. We can then derive the average optimal stopping time as:
\beq
 n^*(c)  = \left[ \sum_{a | e_{\rm g} \le e_a \le C^*(c)} \mathcal P(e_a) \right]^{-1}\,.
\label{eq:meanstoppingtime}
\eeq
The optimal energy $E^*$ can can also be derived using $C^* = E^* + n^* c t_{\run}$.

It is useful to note that the optimal cost $C^*$ reduces to two well-known benchmarking quantities in the small and  large $c$ limits. In the small $c$ limit, the time-to-solution ${\rm TtS}$  can be recovered as follows:
\beq
{\rm TtS} = (C^*(c) -  e_{\rm g})/c, \quad c\rightarrow 0\,.
\label{eq:tts}
\eeq
In the large $c$ limit, the average energy $\bar E$ can be recovered as follows:
\beq
\bar E  = C^*(c)-c t_{\run} , \quad c\rightarrow \infty\,.
\label{eq:en}
\eeq

Next we examine VQA as an algorithm for solving classical binary optimization problems in the form:
\beq
E(z_i) = \sum_{r<s}^N J_{ij} s_i^r s_i^s, \quad s_i^r =  2 z_i^r-1 =   \pm1\,.
\eeq
 We have considered VQA circuits of the following type:
 \beq
 |\Psi^{\rm out}\rangle =\prod_{d}^D U_{d}(\gamma^d_r,\chi^d_r,\zeta^d_{rs}) |\Psi^{\rm in}\rangle\,,
 \eeq
where 
\bea
 U_{d}(\gamma^d_r,\chi^d_r,\zeta^d_{rs}) &= & \exp\left({-i\sum_r^N \gamma^d_r \sigma^y_r}\right) \exp\left({-i\sum_r^N \chi^d_r \sigma^x_r}\right) \nonumber \\
&& \exp\left({-i\sum_{r<s} \zeta^d_{rs} J_{rs}\sigma^z_r \sigma^z_s}\right)\,.
\eea
 The parameters $0< \gamma^d_r,\chi^d_r,\zeta^d_{rs} < 2\pi$ are the phases to be optimized by the VQA algorithm. We have considered $D=3$ in our experiments. The circuit has been initialized with
\beq
|\Psi^{\rm in}\rangle = \prod_i \frac{1}{\sqrt{2}}(\ket{1} + \ket{0})\,,
\eeq
and the initial phases of all gates are randomly picked.

 We have considered 3 ensembles of 250 fully connected  optimization problems of the form $H_{\rm P}$ with $N= 6, 8, 10$ variables and randomly generated $J_{ij} \in \{\pm1,\pm2,\dots,\pm 9,\pm10 \}$. For each instance, we have run the VQA algorithm 100 times, with random initializations of the phases of the circuit\footnote{The $N=10$ ensemble includes only 10 random initializations to limit the utilization of computational resources needed to simulate the VQA circuit. }. All 100 VQA runs are independent. At each step $n$, or iteration of the classical routine, we have computed:
\beq
\mathcal P_n(e_a)  = \frac{1}{100}\sum_{g=1}^{100} P_{n,g}(e_a)\,,
\eeq
where we have used $g$ to indicate the repetition index that we have used to average over several random initializations of the VQA circuit. Finally we use $\mathcal P_n(e_a)  $ to compute $C^*_{n}(c)$, as a function of $c$ and $n$, using Eq.~\ref{eq:opt_E}. Notice that Eq.~\ref{eq:opt_E} depends on $t_{\run}$. In the case of the VQA, we assume $t_{\run}$ to be proportional to the number of circuit evaluations  required by the classical routine to update the circuit phases. We thus consider the following expression for the computational cost:
\beq
n c t_{\run} \rightarrow c n_{\rm evals}\,.
\eeq
\begin{figure*}[ht]
\begin{center}
\subfigure[\, ]{\includegraphics[width=0.36\textwidth]{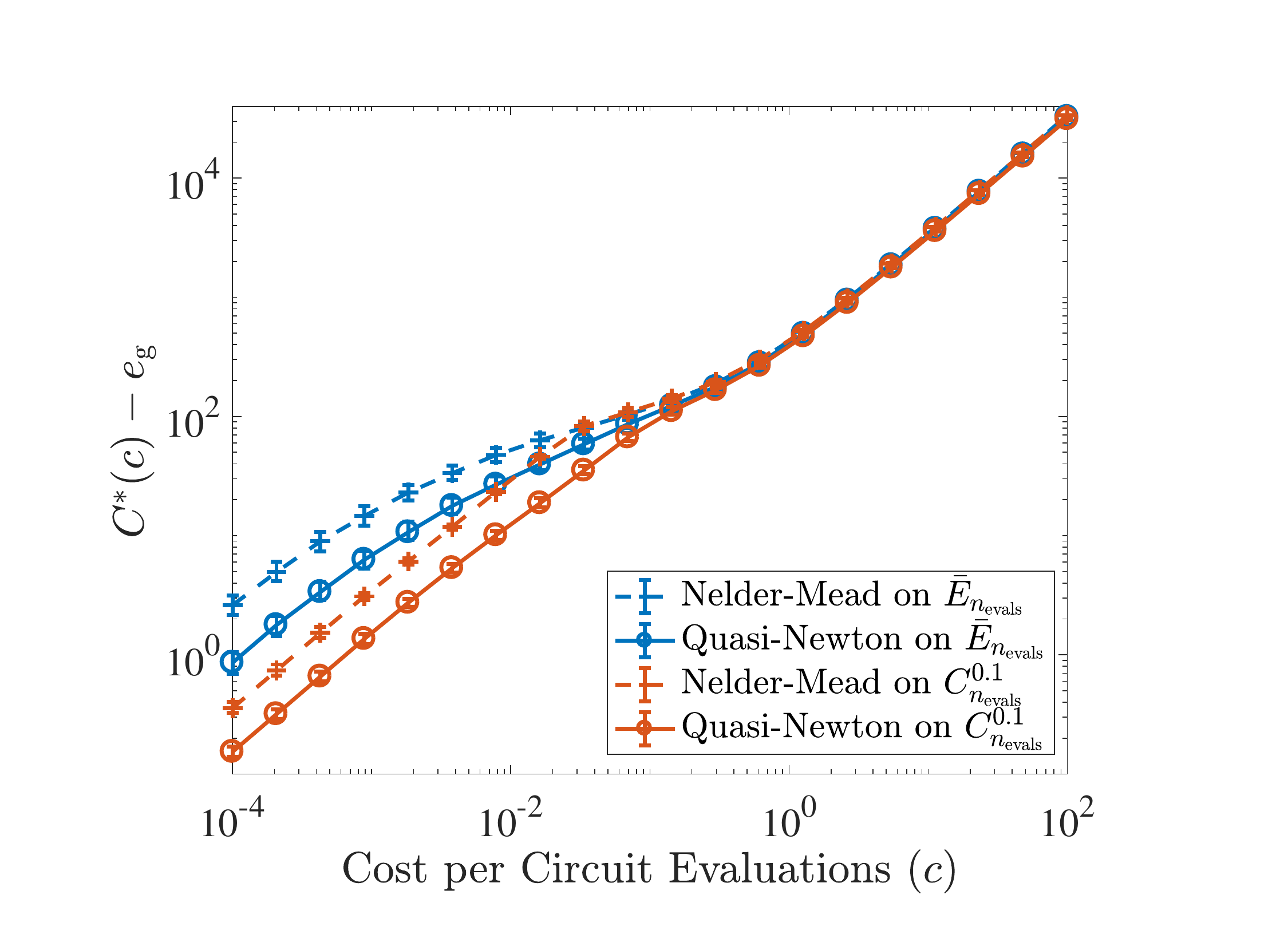}\label{fig:2a}}\hspace{-1cm}
\subfigure[\, ]{\includegraphics[width=0.36\textwidth]{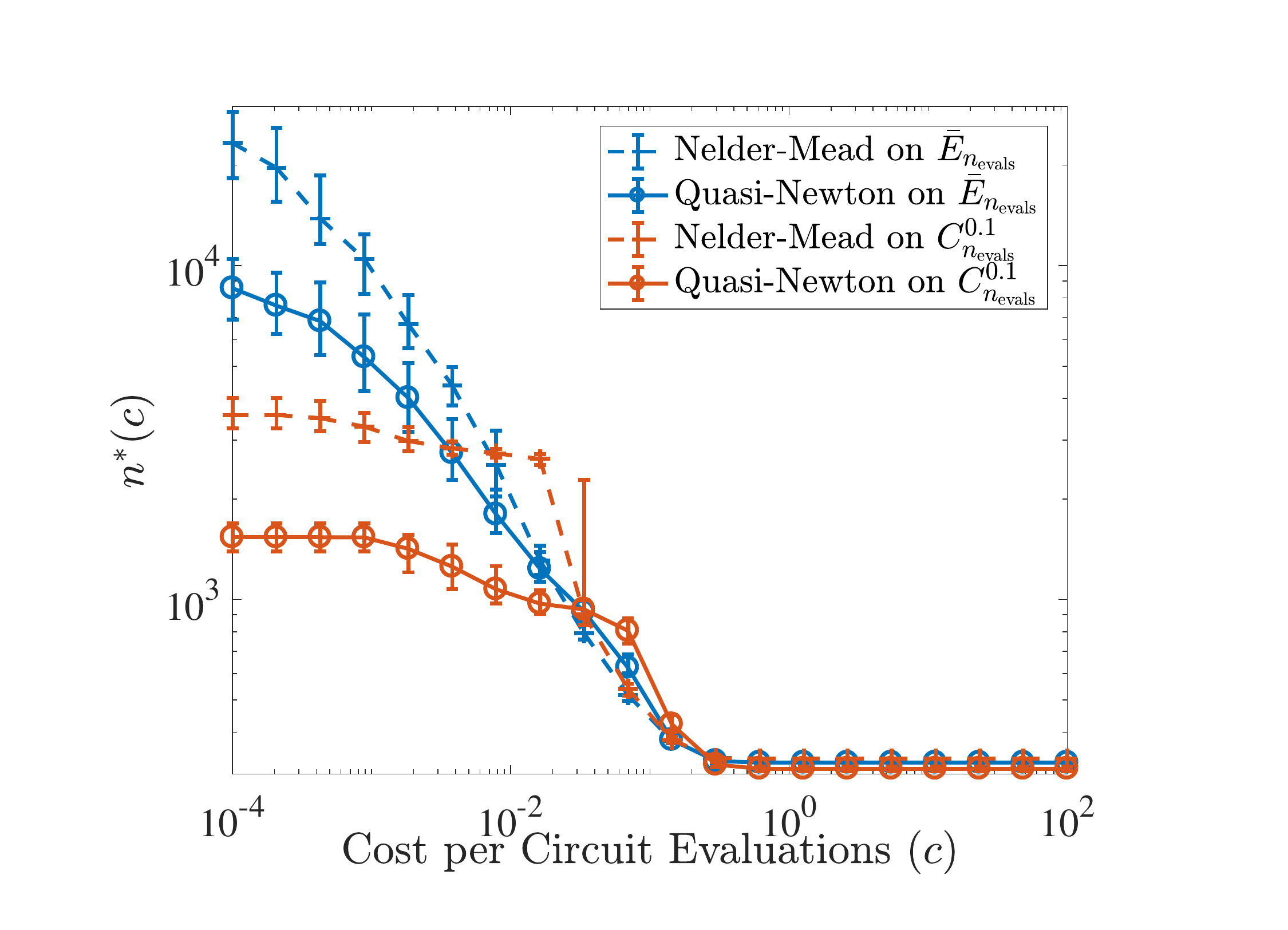}\label{fig:2b}}\hspace{-1cm}
\subfigure[\, ]{\includegraphics[width=0.36\textwidth]{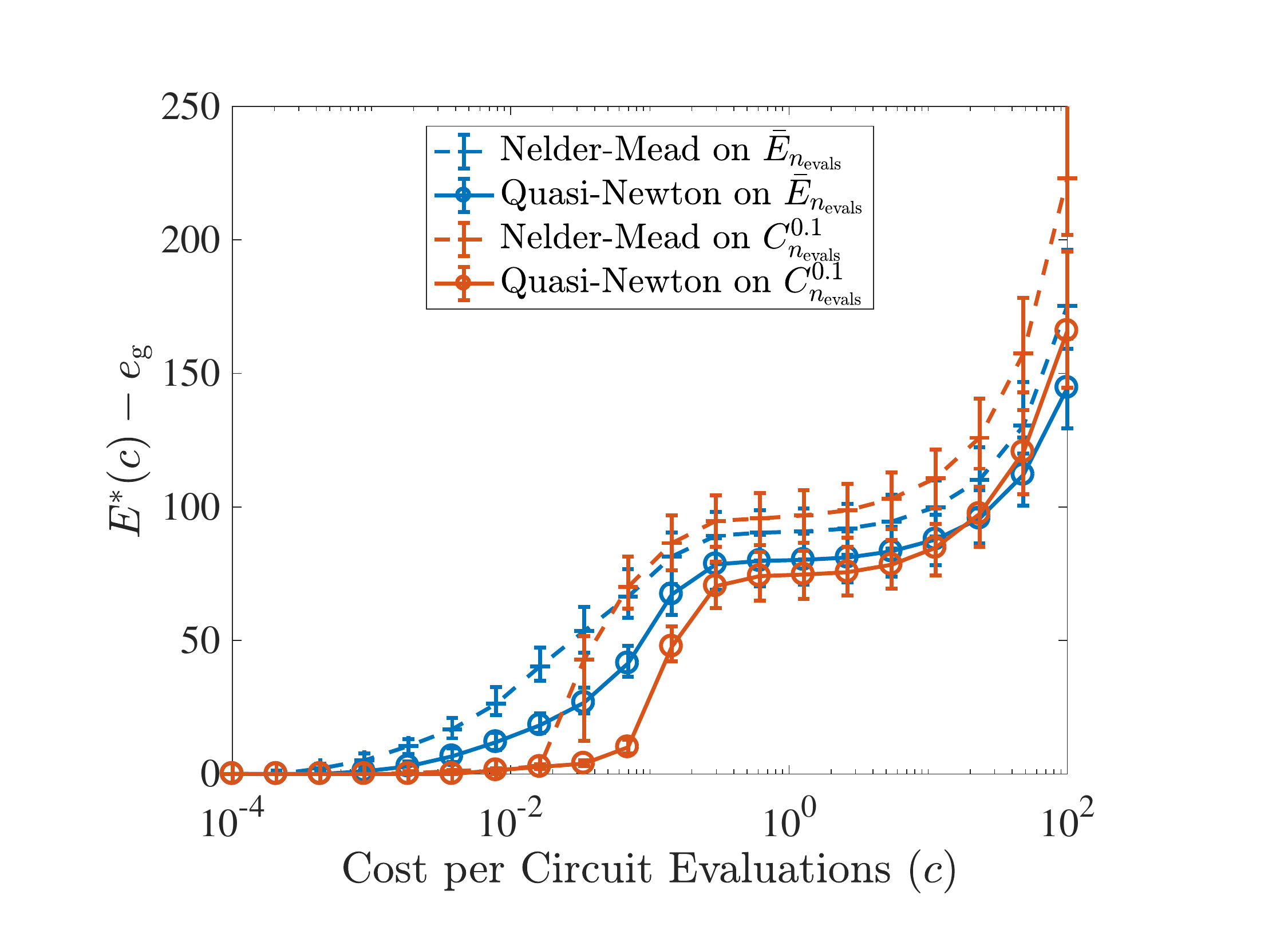}\label{fig:2c}}
\caption{Performance of VQA (depth-3 quantum circuit) on a randomly generated ensemble of 100 QUBO problems with $N=8$ as a function of the cost $c$, after optimizing the performance of   $\tt fminsearch$  and  $\tt fminunc$ in terms of ${n_{\rm evals}}$. (a) Optimal cost $C^*(c)$. (b) Optimal number of circuit evaluations  $n^*(c)$. (c) Optimal energy $E^*(c)$.}
\label{fig:2}
\end{center}
\end{figure*}
\section{VQA for QUBO}
\label{sec:NUM}

We benchmark two standard optimizers included in Matlab. $\tt fminsearch$ uses a Nelder-Mead simplex algorithm and  $\tt fminunc$ is a quasi-Newton method. As mentioned before, both algorithms are applied to a VQA circuit with a randomly initialized starting point. Notice that typically each iteration of $\tt fminsearch$ uses one function evaluation while $\tt fminunc$ uses a number of function evaluations proportional to the number of parameters to optimize.

We have considered two different cost functions to be evaluated and minimized by the two algorithms. First, we have used the average energy:
\beq
\bar E_{n_{\rm evals}} \equiv  \sum_{a} e_a \mathcal P_{n_{\rm evals}}(e_a)\,.
\eeq
The average energy is the commonly used choice in applications VQA. The use of average energy, which is also the expectation value of the problem Hamiltonian, is mostly inspired by quantum chemistry, in which evaluating the minimum of tis quantity is the purpose of the whole algorithm. We question such a choice when VQA is used to solve classical optimization problems. In such case, the goal of the algorithm is not to minimize $\bar E_{n_{\rm evals}}$, but rather to maximize the probability to find solutions that have the smallest energy $e$ possible. We thus consider the following quantity too:
\beq
C^{0.1}_{n_{\rm evals}}  \equiv C^*_{n_{\rm evals}}(c)\,, \quad c=0.1\,.
\eeq
The optimal cost $C^{0.1}_{n_{\rm evals}}$ computed at a small, but not too small ($c=0.1$), value of $c$ weighs the lower tail of $\mathcal P_{n_{\rm evals}}(e_a)$ without being too weighted on the ground state. This choice helps the solver to smoothly increase the weight of the output wave-function $ \Psi^{\rm out}$ on low-energy states.

Numerical results for the case $N=8$ are shown in Fig.~\ref{fig:1} and~Fig.~\ref{fig:2}. For each instance, we have computed $C^*_{n_{\rm evals}}(c = 0.02)-e_{\rm g}$. This quantity, shown in Fig.~\ref{fig:1a}, is more convenient to plot since we always have $C^*(c)  > e_{\rm g}$. Figure~\ref{fig:1a} (as well as all subsequent figures) shows the median value, and the error bars the 25th and 75th percentiles for the 250 instances.  Notice that $C^*_{n_{\rm evals}}(c)$ has a minimum at a certain number of circuit evaluations ${n_{\rm evals}}$. As usual, any solver has an optimal number of iterations. This is due to the fact that there is a initial number of evaluations in which the classical routines are effective in reducing $C^*_{n_{\rm evals}}(c)$ despite increasing the length of the calculations and thus increasing the computational cost $c n^*_{n_{\rm evals}}(c)$. After some time, the classical routines are less effective in optimizing the circuit to increase the probability to obtain good quality solutions, and  $C^*_{n_{\rm evals}}(c)$ starts growing like $c n_{\rm evals}$ due to the increase in the computational effort. Notice that $\tt fminunc$, as expected, has faster convergence and usually allows to obtain a smaller  $C^*_{n_{\rm evals}}(c)$.

\begin{figure*}[ht]
\begin{center}
\subfigure[\,  ]{\includegraphics[width=0.36\textwidth]{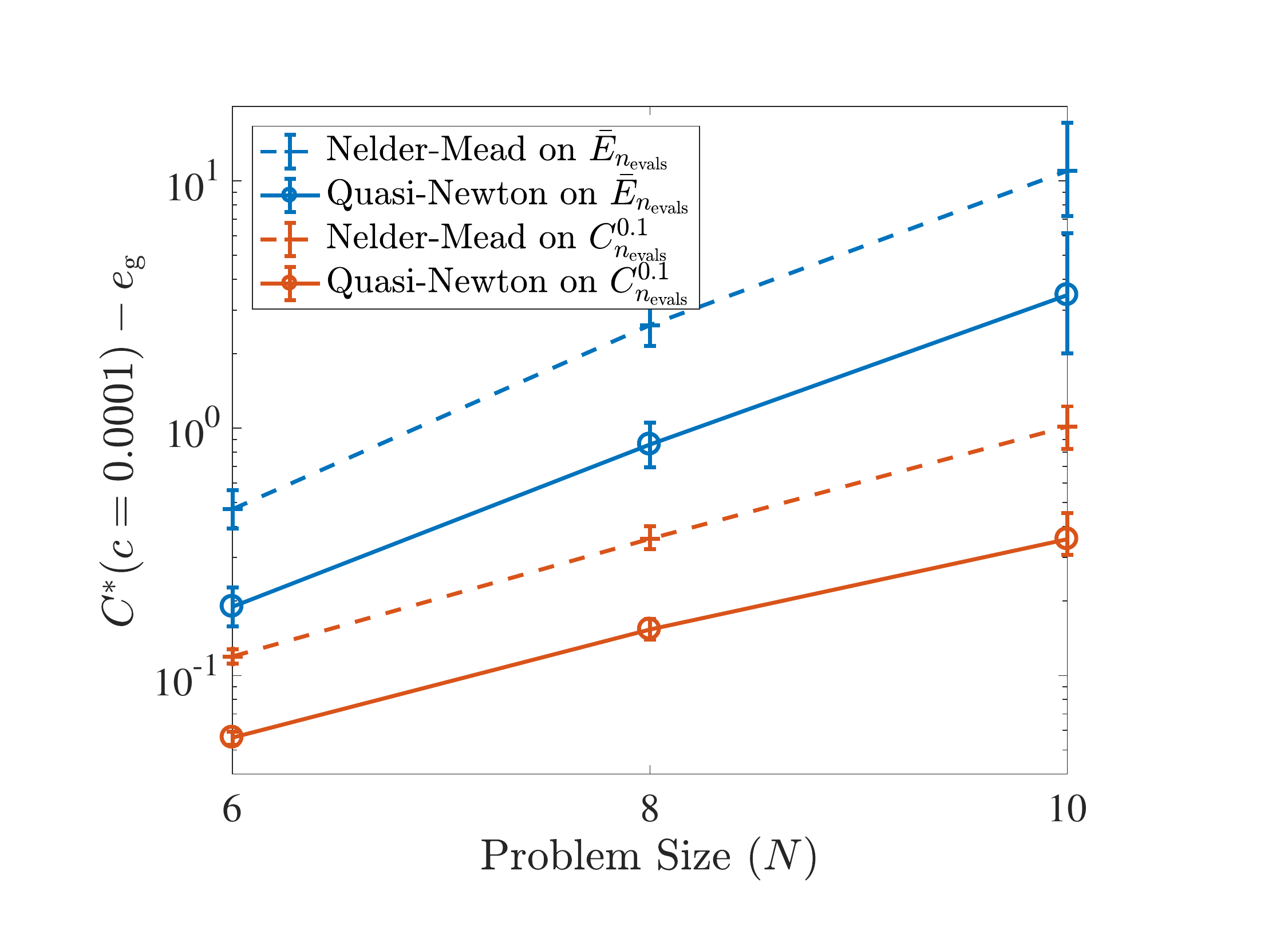}\label{fig:3a}}\hspace{-1cm}
\subfigure[\,  ]{\includegraphics[width=0.36\textwidth]{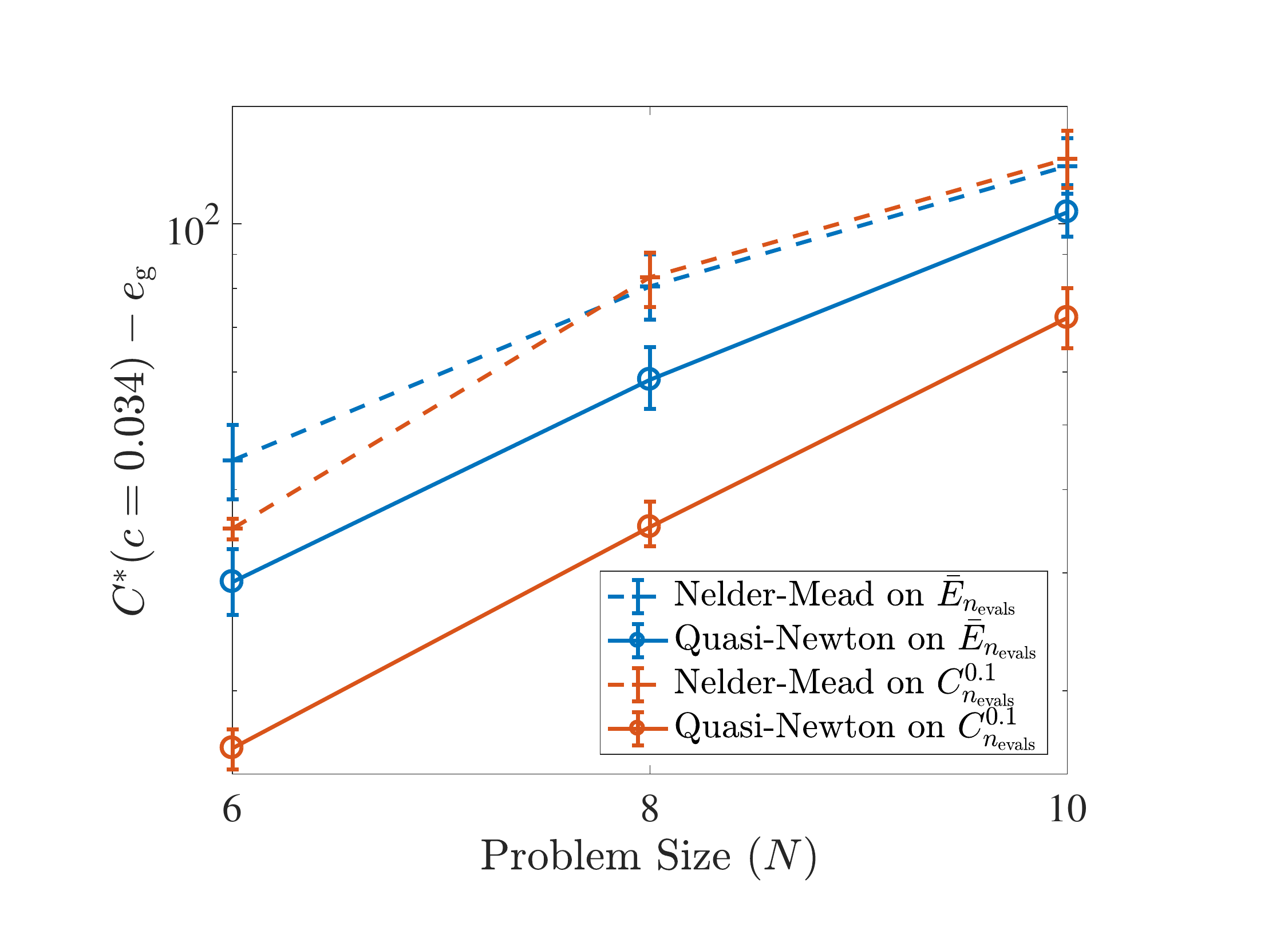}\label{fig:3b}}\hspace{-1cm}
\subfigure[\,]{\includegraphics[width=0.36\textwidth]{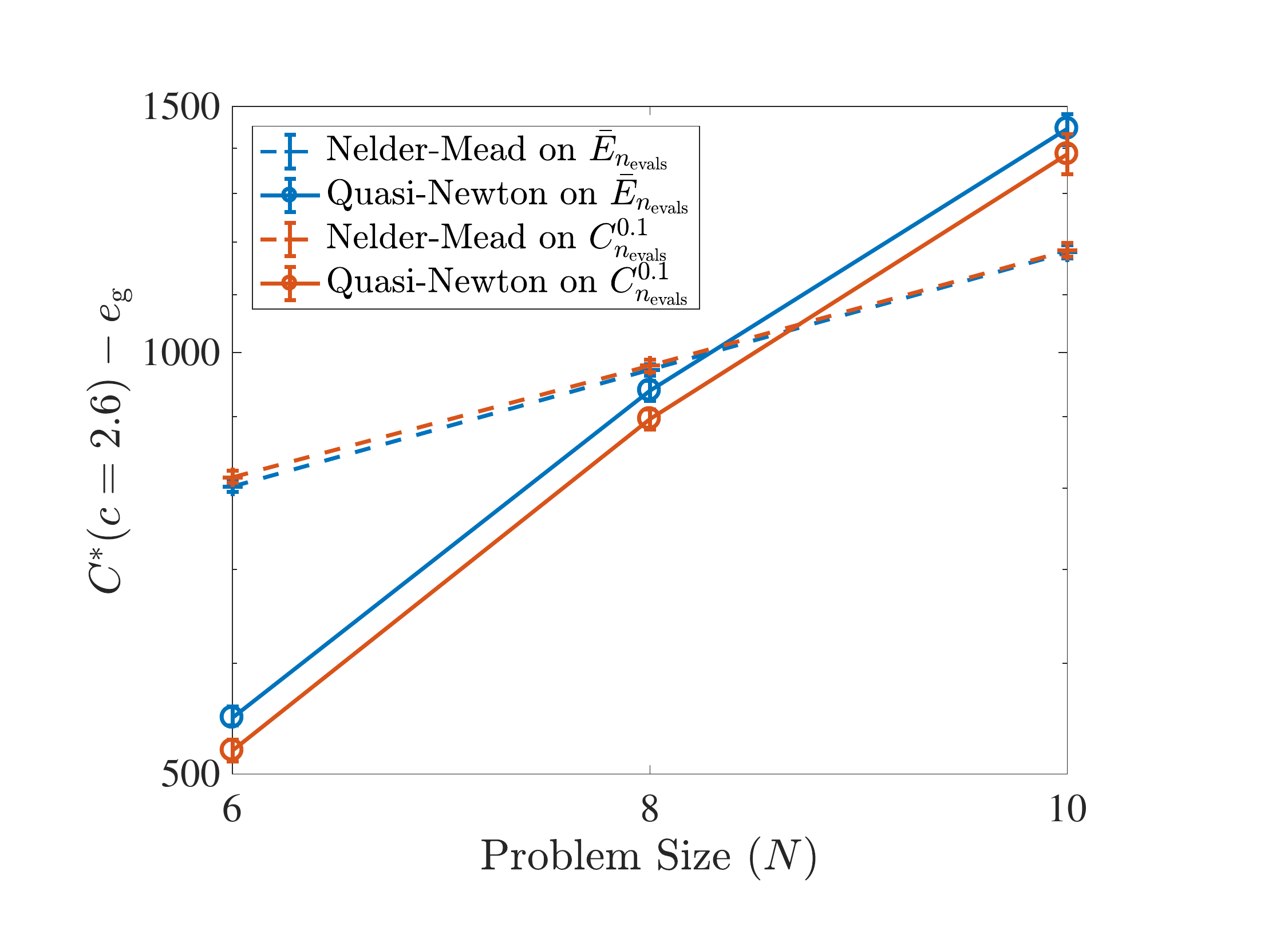}\label{fig:3c}}
\caption{Scaling of $C^*(c)$ as a function of the problem size for $N=6$, $N=8$ and $N=10$. (a) $c=0.0001$. (b) $c=0.034$. (c) $c=2.6$. }
\label{fig:3}
\end{center}
\end{figure*}

As we said before, for small $c$, and in particular as long as $C^*_{n_{\rm evals}}(c)- e_{\rm g}< e_{\rm f}-e_{\rm g}$ ($e_{\rm f}$ is the energy of the first excited state), $C^*_{n_{\rm evals}}(c)$ is proportional to the ${\rm TtS}$. The y-axis of  Fig.~\ref{fig:1a} can also be considered thus as showing ${\rm TtS}$ for the VQA. Indeed, the behavior of  $C^*_{n_{\rm evals}}(c)- e_{\rm g}$ is qualitatively very similar to the optimal number of circuit evaluations $n^*_{n_{\rm evals}}$ (obtained according to Eq.~\ref{eq:meanstoppingtime}), which is shown in Fig.~\ref{fig:1b}. Both Figs.~\ref{fig:1a} and ~\ref{fig:1b} show that for small $c$ the choice of  $C^{0.1}_{n_{\rm evals}}$ is much more effective than $\bar E_{n_{\rm evals}}$. The reason for this is clearly shown in Fig.~\ref{fig:1c}, were we plot the shifted expected energy $E^*_{n_{\rm evals}}(c) - e_{\rm g}$. We see that using $C^{0.1}_{n_{\rm evals}}$ allows the VQA circuit to output a state $ \Psi^{\rm out}$ that quickly converge to a (superposition of) ground state(s). On the other hand using $\bar E_{n_{\rm evals}} $ pushes the circuit to reduce $\bar E_{n_{\rm evals}} $ but at the same time typically reduces the weight of the output state $ \Psi^{\rm out}$ on low-energy states. This effectively reduces the performance of VQA as an optimizer, in which the goal is to obtain states of the computational basis with low energy. We observe that this happens more efficiently when we optimize $C^{0.1}_{n_{\rm evals}}$.

 As customary, we must optimize the classical solvers in terms of the number of circuit evaluations to minimize  $C^*_{n_{\rm evals}}(c)$:
 \beq
  C^*(c) =  \min_{n_{\rm evals}}C^*_{n_{\rm evals}}(c)\,.
  \label{eq:optimize_n}
 \eeq
The quantity above is plotted in Fig.~\ref{fig:2a}, for the $n=8$ ensemble. At both the left and right extremes of the x-axis we see the linear in $c$ behavior of $C^*(c)$ that is related to  ${\rm TtS}$ and $\bar E$ (see Eqs.~\ref{eq:tts} and~\ref{eq:en}). At intermediate values of $c$ there is a transition region in which $C^*(c)$ represents a non-trivial balance between solution quality and computational effort. In Fig.~\ref{fig:2} we see that using $C^{0.1}_{n_{\rm evals}} $ is more effective than $\bar E_{n_{\rm evals}}$ over the whole range of $c$ values. It is likely that the best cost function for the classical routine is $C^*_{n_{\rm evals}}(c)$ itself, with $c$ not fixed to $c=0.1$, as in our experiments, but equal to the actual value of $c$. We have not tried this numerically since it requires performing independent VQA optimizations at each value of $c$, a very intensive computational task. In Fig.~\ref{fig:2b} we show the optimal number of circuit evaluations $n^*(c)$, which is proportional to the optimal duration of the optimization process. Notice that $n^*(c)$ is typically achieved by repeated (and independent) runs of the optimization routines after the number of iterations $n_{\rm evals}$ is optimized as in Eq.~\ref{eq:optimize_n}. The quantity $n^*(c)$ is a monotonically decreasing function of $c$. This is intuitive: the optimal duration of the computation becomes smaller as the cost of the computation per circuit evaluation $c$ increases. As we can see in Fig.~\ref{fig:2c}, a consequence of this is that the optimal energy $E^*(c)$ grows as a function of $c$: as the computational effort grows with $c$, it is optimal to stop earlier and accept solutions with larger energy.

Finally, we show in Fig.~\ref{fig:3} how $C^*(c)$ scales as a function of the problem size at three different values of $c$. We have considered three different problem sizes: $N=6$, $N=8$ and $N=10$. At small $c$,  $C^*(c)$ is equivalent to a ${\rm TtS}$ and should scale exponentially. A conclusive scaling analysis cannot be conducted with such small problem sizes. However, Fig.~\ref{fig:3a} is a strong indication that a better choice of the cost function for the classical routine (in our case $C^{0.1}_{n_{\rm evals}}$ versus $\bar E_{n_{\rm evals}} $) can improve the scaling of VQA in solving classical optimization problems at small $c$. At intermediate values of $c$, see Fig.~\ref{fig:3a}, $C^*(c)$ doesn't show clear scaling behavior since $C^*(c)$ is a complicated balance between solution quality and computational time. At large values of $c$, one should observe a polynomial scaling of  $C^*(c)$~\cite{vinci2016optimally}. Figure~\ref{fig:3c} shows that  $\tt fminsearch$ may scale better than  $\tt fminunc$ at large $c$, an opposite to what happens for small $c$. This is due to the fact that $\tt fminunc$ typically requires a larger number of circuit evaluations per iteration that grows with the problem size. This makes $\tt fminunc$ more costly than $\tt fminsearch$ at large values of $c$. This is again another consequence of using optimal costs for benchmarking. The choice of solver usually depends on the value of $c$, which is typically determined by practical considerations.

\section{Discussion and Outlook}
\label{sec:DIS}

In this manuscript, we proposed optimal stopping as a benchmarking approach to asses the performance of VQA algorithms for solving classical QUBO problems. Arguably, any complete evaluation of VQA should quantify the outer-loop optimization time since VQA is inherently a hybrid classical-quantum algorithm. The cost of the classical part is determined by the optimization method used to search for optimal quantum evolution parameters which consequently determines the scalability and competitiveness of VQA. The optimal stopping approach explicitly and elegantly encodes in one single quantity - the optimal cost - both the quality and the computational cost of the optimization process. The optimal stopping method introduced in this work is a way to accurately assess the power of VQAs and to determine the best classical routines for the optimization of the VQA quantum circuit. 

In many situations, like in quantum chemistry, the VQA circuit is thought of as an algorithm to generate a variational approximation of the quantum state that minimizes the expectation of a final quantum Hamiltonian. When this is the case, the classical optimization routines must adjust the parameters of the quantum circuit to lower such expectation. On the other hand, when the VQA circuit is used to solve classical QUBO problems, the final quantum state provides a superposition of (low energy) classical states that must be extracted after measurements in the computational basis. In this situation, the choice of the cost function for the classical routine is arbitrary. With an appropriate choice, the classical optimization routine should adjust the parameters of the quantum circuit to generate a quantum state with larger weights on low-energy classical states. Indeed, we have found that using cost functions for the classical routine that are more sensitive to low-energy classical states than the expectation value of the problem Hamiltonian, we can significantly improve the performance of the VQA algorithm and even improving it's scaling behavior.

We hope this study further motivates developing novel scalable algorithms for classical optimization in VQAs.

\end{document}